# Modeling the Semantics of States and State Machines

Sabah Al-Fedaghi

*Department of Computer Engineering, Kuwait University, Kuwait*

**Abstract:** A system's behavior is typically specified through models such as state diagrams that describe how the system should behave. According to researchers, it is not clear what a state actually represents regarding the system to be modeled. Standards do not provide adequate definitions of or sufficient guidance on the use of states. Studies show these inconsistencies can lead to poor or incomplete specifications, which in turn could result in project delays or increase the cost of the system design. This paper aims to establish a precise definition of the notion of states and state machines, a goal motivated by system modelers' (e.g., requirement engineers') need to understand key concepts and vocabulary such as states and state machine, which are major behavioral modeling tools (e.g., in UML). "State" is the main notion of a state machine in which events drive state changes. This raises questions about the nature of these state-related notations. The semantics of these concepts is based on a new modeling methodology called the thinging machine applied to a number of examples of existing models. The thinging machine semantics is founded on five elementary actions that divide the static model into changes/states upon which events are defined.

## Introduction

Abstraction is one of the most important tools used in computer science (Lewis and Lacher, 2016). The design and implementation of complex systems cannot be done without it (Van Otterlo, 2009). Many forms of abstraction are used in modeling systems and among them state abstraction is one of the most common. For example, in chess a complete configuration of black and white pieces on a board is the chess machine's state (Van Otterlo, 2009). State machines (abstract model consisting of states, input and mapping of input to states) typically specify how a system should behave. These notions are essential ingredients in engineering systems e.g., the (Defence Materiel Organisation, 2011) Development Guide requires the identification of all of the applicable states for the solution-classes (Olver and Ryan, 2014). States provide means to "identify different sets of performance requirements for different sets of conditions that will be encountered by the system" (Space and Missile Systems Center, 2005). State-centric specifications not only serve to identify what is to be accomplished but also specify how to design the system (Wasson, 2005).

*Problem*

According to (Olver and Ryan, 2014), the various methodologies for specifying states do not provide a consistent message or framework of what constitutes a state. They emphasized, "The wide variation of definitions [of system states] demonstrates that no consistent structure exists." For example, the US Department of Defense military standards do not adequately define or provide sufficient guidance in the use of states (Olver and Ryan, 2014) and the INCOSE Handbook (INCOSE-TP-2003-002-03.2, 2010) does not provide a definition for states and modes (Space and Missile Systems Center, 2005). The (IEEE 610.12-1990, 1990) standards confuse the issue even more by defining state as "a condition or mode of existence that a system, component or simulation may be in" (Olver and Ryan, 2014). These inconsistencies can lead to poor or incomplete specifications, which in turn could result in project delays or increased cost of system design. While the notion of state is used to describe a system, "there is little guidance as to what constitutes a state, which is made worse by poor usage of the terms" (Olver and Ryan, 2014). Although there are many different ways in which state-based specifications can be represented,

"there are few detailed methodologies that provide a sound development framework to assist the organization to develop the state based specification" (Olver and Ryan, 2014). In the Unified Modeling Language (UML; Object Management Group, 2015) and hence the System Modeling Language (SysML; Object Management Group, 2017), "It is not clear what a state actually represents regarding the system to be modeled…. Aside from UML, there are 'state' elements or concepts used in different languages, tools and methods with different semantics" (Baduel et al., 2018). In academia, the scarce research investigating the different ways in which states are defined and used within the systems engineering disciplines (Olver and Ryan, 2014) does not include many references (Edwards, 2003; Wasson, 2011).

A Finite State Machine (FSM) is used as a behavioral model that characterizes behavior using transitions. A basic FSM defines a set of states, a set of events it responds to and a set of transitions that map a state and event to (next) state. FSMs are a major behavioral modeling tool and part of UML, which is the current main information systems modeling notation. UML 2 adopted the Statechart model, which is an extension of the basic FSM model. In UML, all objects have a state. An object either does or does not exist. If it exists, then it has a value for each of its attributes. Each possible assignment of values to attributes is a state. According to (Olver and Ryan, 2014), "The clarity and unambiguous [nature] of the [UML] FSM model (if done right) is by itself a sufficient justification for using it." Nevertheless, the implementation of FSMs in Object-Oriented (OO) languages "often suffers from maintenance problems" (van Gurp and Bosch, 1999). In general, according to (Wagner et al., 2006),

> In the scientific world the definition of the state machine dominates in software application. Discussing hardware design, scientific papers concentrate on model definitions, optimization of state number and verification methods. All those theoretical topics are of little practical usefulness and do not make too much sense in the design of an industrial control system. Hence, *the knowledge and the use of state machines in industry are half-hearted and accompanied by several misunderstandings due to lack of a sound theoretical basis*. (Italics added).

*Aim*

This paper aims to further the understanding of the semantics (interpretation of meaning) of states and state machines. FSM semantics is severely complicated because formalization takes multiple approaches and several state machine dialects currently exist, each subtly different from the others (Crane, 2006). OMG has issued specifications for UML-based semantics, behavioral semantics, static semantics and structural semantics (Object Management Group, 2019). Behavioral semantics provide "the denotational mapping of appropriate language elements to a specification of a dynamic behavior resulting in changes over time to instances in the semantic domain about which the language is making statements" (Object Management Group, 2007). In this study, semantics refers to the meaning of individual constructs (Harel and Rumpe, 2004), such as FSM notions of state, event, change and machine. The interest in semantics is motivated by the need of system modelers (e.g., requirement engineers) to know key concepts and vocabulary of major behavioral modeling tools (e.g., in UML and SysML).

*Focus on Semantics of Individual Constructs*

Hatley and Pirbhai (2013) astutely pointed out that "theoreticians enjoy long philosophical discourses on the exact meaning of state." In physics, a state is usually defined as a system's complete set of possessed properties e.g., mass or electric charge; (Craig, 1998). Olver and Ryan (2014; Baduel et al., 2018) provide comparisons of the definitions of states given within various articles in the modeling literature. A sample of the definitions of state is as follows:

- A static snapshot of the set of metrics or variables needed to fully describe the system's capabilities to perform the system's function (Baduel et al., 2018)
- The overall condition of a system or a subsystem, whereby events may be related hierarchically or exclusively (Edwards, 2003)
- The operational or operating condition of a system of interest required to safely conduct or continue its mission (i.e., "state of operation;" Wasson, 2005)
- The events happening within the system at any point in time (Holt, 2004)
- An exact operating condition of a system (Jenney et al., 2010)
- A mode or condition of existence for a process or other system component, as determined by current circumstances (Hoffer et al., 1996)

Baduel et al. (2018) cross-referenced the various definitions and established a few characteristics of states:

- States characterize a thing (e.g., a system).
- States relate to a specific kind of information, operation, readiness, energy, etc.
- States are evaluated or considered at a given time.

For (Balabko and Wegmann, 2003), the state of an object at an instant in time is the condition of the object that determines the set of all sequences of actions in which the object can take part. Action is something that happens. Thus, the concept of state is intertwined with the concept of action and these modeling concepts cannot be considered separately.

Olver and Ryan's (2014) survey of existing methods of state presentation included state charts (Harel, 2007), phases (re-mission, mission and post mission; (Wasson, 2005), flow graphs (Van Court Hare and Starr, 1967), UML state machines (OMG, 2007) and temporal logic modelling (Hoffer *et al.*, 1996). According to (Baduel *et al.*, 2018), the state definition in UML cannot be used as a definition of the concept because "to represent the behavior, we have to define elements characterizing the execution of capabilities and not the information used to allow it, which seems to be the goal of the state machines in UML."

It is difficult to say what an *event* is in the context of FSM when reading the literature. The notion of an event is understood in general from many examples as a (interior or exterior) trigger of actions in a machine. Most works in FSM literature do not define the term *event* directly. For example, the valuable work in (Avnur, 2015) analyzes many interesting aspects of the FSM. It includes more than 100 literary occurrences of the term *event*. The first appearance of the word *event* is on the second page: "The dynamic part of that behavior definition is done by defining the transitions from state to state-when it occurs, i.e., what event causes it." Then there are numerous mentions of the word event, which include the following quotes:

- Integrity constraints … that are important or *event* critical to the application
- *Events* are applied or "signaled" to the FSM through its interface
- An *event* transitions the system from various states to a common state
- List of events and their meaning: Delete: Delete the record, Edit: Prepare the record for editing, etc

In the technical literature of state machines using models such as UML and SysML, the states are the stages or situations during a system's life when it performs some activity or waits for some event (Engel, 2010). In this description, an *event* is "an incident provoking (or not) a reaction of the state machine [and] a transition is a specification of how a state machine reacts to an event. It specifies a source state, the event triggering the transition, the target state, guard and actions" (Henry, 2010). An *action* is an operation executed during the triggering of the transition. A *guard* is a Boolean operation that is able to prevent the triggering of a transition that would otherwise fire.

Of course, the mathematical definition of FSM based on automata theory provides a clear definition of an event as an (user or system) input that triggers a change from one state to another in the FSM. In this study, we define events in terms of changes which, in turn, are defined in terms of five elementary actions.

*Semantics and FSM*

According to (Yan, 2017), FSM is an approach to bridging the gap between the real world and semantic space by using events. Henry (2010) articulates that state machines are "the description of a thing's lifeline" that includes different stages. The focus in this study is on semantics of modeling; that is, the "meanings" of constructs is a central concern in the paper. This 'meaning' will be defined in terms of five basic actions.

Take for example the FSM presented in (Clayton, 2018) that involves designing a shopping cart encapsulated in an Order entity. The initial state is 'creating'; in this state, the customer adds items to the order. When the customer is done, they will 'check out,' causing the order model to transition to 'finalizing', where no more items can be added to the cart. The interest of this paper is the *meaning* of creating, finalizing, paying, etc. According to (Engel, 2010), these terms represent stages during the system's life when it performs some activity or waits for some events. As will be clear at the end of the analysis in this study, the FSM is a shorthand notation to a richer conceptual representation called thing machine.

*Approach: FSM and Thinging Machine*

We will analyze two abstract machines: (a) An FSM will be "stretched" (activities of states will be added) to construct (b) an abstract machine called a Thinging Machine (TM). The TM will be the mechanism used to analyze and understand FSM. Such a study gives us an opportunity to further understand the TM model and demonstrate its viability in the conceptual modeling field.

*Structure of the Paper*

The next section presents an enhanced review of the TM that has previously been used in several papers (Al-Fedaghi and Al-Fadhli, 2020; Al-Fedaghi and Behbehani, 2020; Al-Fedaghi and Haidar, 2020; Al-Fedaghi and Al-Otaibi, 2019; Al-Fedaghi and Haidar, 2019). The remaining sections describe how the TM can be applied to FSMs.

## Thinging Machines

In the state machine paradigm, the categories of modeling are states, events and transactions. A TM is built from things and machines. Things emerge from their machine; for example, as a set (thing) that manifests the process of "grouping" its members (machine). The philosophical foundation of this approach is based on Heidegger's notion of thinging (Heidegger, 1975). Riemer *et al.* (2013) stated that Heidegger's philosophy gives an alternative analysis of eliciting knowledge of routine activities, capturing knowledge from domain experts and representing organizational reality in authentic ways (Riemer *et al.*, 2013). In TM modeling, things are unified with the concept of a process by being viewed as single ontological things/machines, or thimacs, which populate the world. A unit in such a universe has a dual role as a thing and as a machine.

The simplest type of a complete thimac Machine (M) is shown in Fig. 1. The machine has five actions (also called stages of M): Create, process, release, transfer and receive. An action is defined as one of these five stages. The arrows in M refer to the *order* ("before" and "after") among the five actions. Flow of things refers to the "conceptual movement" of things (e.g., among stages). Note the strong thesis in TM modeling that all actions in the world are categorized as one of the five M actions.

The stages of M can be described as follows.

*Arrive*

A thing flows to a new machine (e.g., an external signal arrives to an FSM).

*Accept*

A thing enters M (the arriving thing is accepted by the event handler in FSM). For simplification, we assume that all arriving things are accepted; hence, we can combine *arrive* and *accept* into the **receiving** stage.

*Release*

A thing is marked as ready to be transferred outside the machine (e.g., in an airport, passengers wait to board after passport clearance).

*Process (change)*

A thing changes its form, but not its identity (e.g., a node in the network machine processes a packet to decide where to forward it).

*Create*

A new thing is born in a machine (e.g., a new shopping cart in the shopping example).

*Transfer*

A thing is inputted into or outputted to/from a machine.

TM modeling includes one additional notation: *Triggering* (denoted by a dashed arrow in this study's figures). Triggering initiates a flow from one machine/submachine to another. Multiple machines can interact with each other through flows or triggering. Triggering is a transformation from one flow to another (e.g., a flow of electricity triggers a flow of air).

The thesis that things are machines and machines are things gives us a tool for handling things as processes. Thus, instead of the notions of class, attributes and methods in object-oriented methodology, TM modeling has processes (machines) and sub processes (submachines).

*Ontology of the TM Model*

TM modeling is based on a category called thimacs (things/machines), which is denoted by ∆ Fig. 2. The ∆ has a dual mode of being: The machine side, denoted as M and the thing side, denoted by T. Thus, ∆ = (*M*, *T*). Machine here acts as a metaphor for a mechanism, or apparatus to represent the "mechanical side" of being. Diagrams in this study will show only the machine side of thimac. According to (Vardi, 2012; Rapaport, 2015), "Turing Machines, the lambda calculus, recursive functions, etc., are all logically equivalent-that these distinct notions are analogous to the wave-particle duality in quantum mechanics: 'An algorithm is both an abstract state machine and a recurs or and neither viewed by itself-' fully describes what an algorithm is. This algorithmic duality seems to be a fundamental principle of computer science."

The claim of the TM model is that ∆ is a fundamental constituent in describing a portion of the world. The model accounts for many thimacs. It is a continuant that has static and dynamic versions of being. The static version is a thing/machine that persists in pretime (to be exemplified later) and undergoes *changes*. The dynamic version also embraces time and hence it embraces events. Ontologically, we construct the thimac as follows:

1. *Actions* definable in terms of five types of actions; e.g., create something
2. *Changes* as chunks of actions
3. Events replace changes when time is inserted in the thimac (e.g., the creation happens at 11:00-12:00 AM)

*Three Levels of Modeling*

According to (Olver and Ryan, 2014), a *static* condition of the system could be defined as a state. A dynamic condition could also be defined as a state, but may better be defined as an operational phase. Such a differentiation could be established by examining the effects of time on the system (Olver and Ryan, 2014). TM modeling presents a clear view of staticity, dynamism and temporality by establishing three levels of representation:

(1) A **static** (atemporal) model, denoted by **S**, is constructed upon the flow of things in five generic actions (activities; i.e., create, process, release, transfer and receive). The **S** model contains all ∆s (thimacs and subthimacs) that are there (no temporality)
(2) A **dynamic** model, denoted by **D**, identifies the changes as subdiagrams of the diagram of S. The **D** model is still an atemporal representation of changes
(3) A **behavioral** model, denoted by **B**, depicts a chronology of events in time. An event is a change that embeds a time subthimac

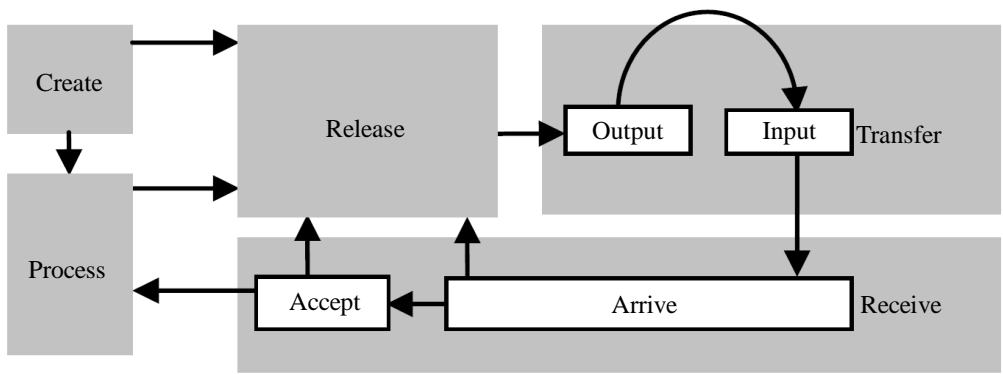

**Fig. 1:** A thinging machine, M

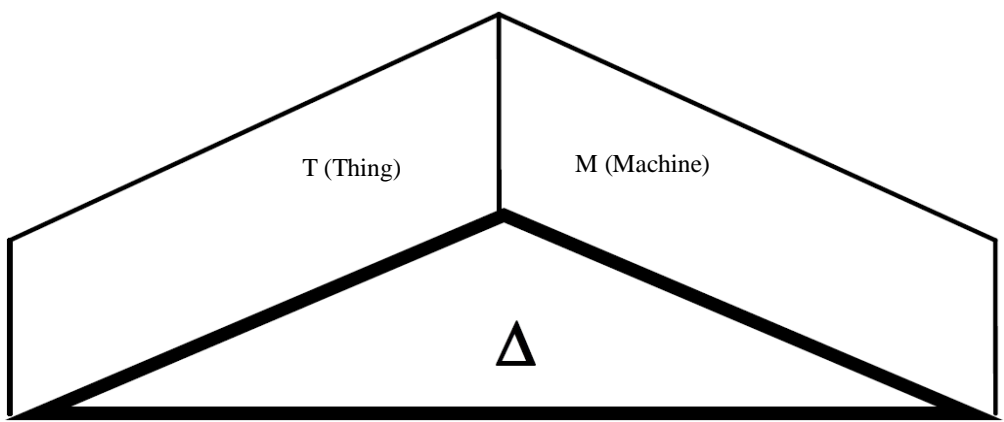

**Fig. 2:** A thimac has a dual mode of being a thing and a machine

## Exemplifying TM Modeling of FSMs

In this section, we provide an example of the method that will be pursued in the remaining part of this paper. The overall paper is an expansion of this method with larger examples and other, more relevant issues. This example is presented by (Tsonev, 2018) and involves an illustration of how a state machine works through modeling a turnstile. The turnstile FSM has a finite number of states (locked and unlocked) with their possible inputs and transitions, Fig. 3.

Figure 4 shows the corresponding **S** model. Initially, the turnstile (circle 1) is locked (2), which prohibits flow through the transfer stage (3). This situation is analogous to a lock and a door (represented by the transfer stage). When the lock is ON, nothing can go through the door.

When a coin is created (generated/appears; 4) and flows to the system (5), it is processed (validated; 6) to unlock (7) the turnstile. This triggers permission to go through the transfer stage (door; 8) when a person (10) pushes through (11) the turnstile. Thus, the person is received in (12), released and transferred (13) outside the turnstile area, which triggers the turnstile to lock (14). Later, we will justify why **S** is called the static model.

Viewing **S** as a collection of small scripts, the modeler's role is to identify consistent scripts that may be constructed from the collection of short scripts.

This results in cutting the **S** diagram into specific subdiagrams (called *events*), producing the **D** model as shown in Fig. 5. The events are:

$E_1$ = The locked turnstile
$E_2$ = No transfer permitted
$E_3$ = A coin inserted and processed
$E_4$ = The turnstile unlocking
$E_5$ = A person pushing through
$E_6$ = Transfer permitted
$E_7$ = A person exiting through the turnstile area to outside

Figure 6 shows the chronology of these events.

In this study, we examine the semantics of the state machine (i.e., states, events, transactions and action?) in terms of TM modeling. The assumption here, as shown in the turnstile example, is that TM modeling is richer and more precise than FSM modeling. Furthermore, the next example clarifies the nature of events and how they relate to the **D** model.

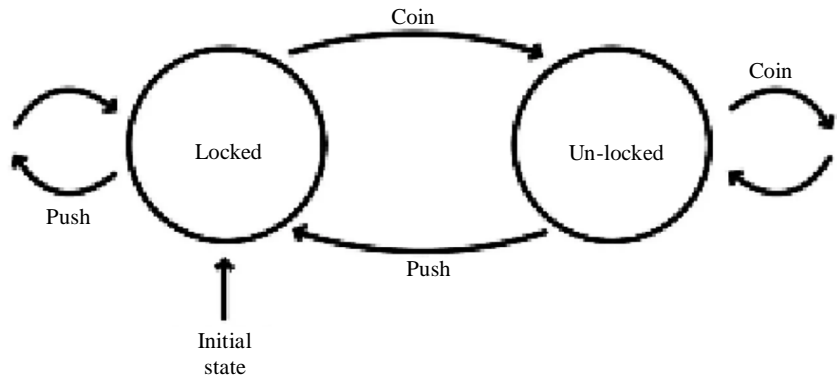

**Fig. 3:** State machine model of a turnstile (adapted from Tsonev, 2018)

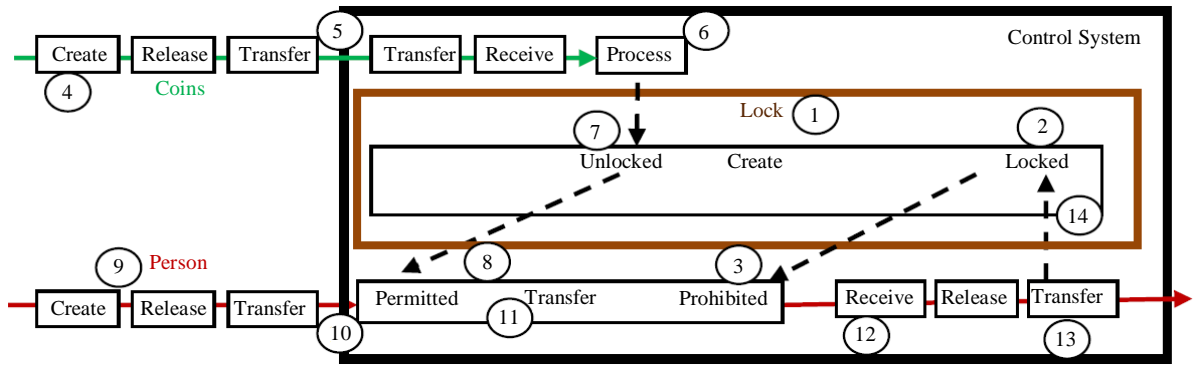

**Fig. 4:** The **S** model of the turnstile

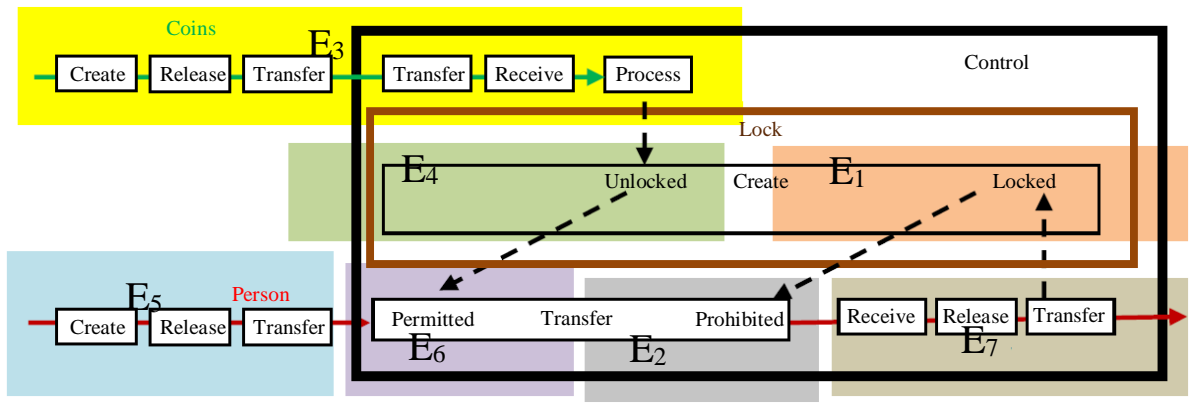

**Fig. 5: D** model of the turnstile

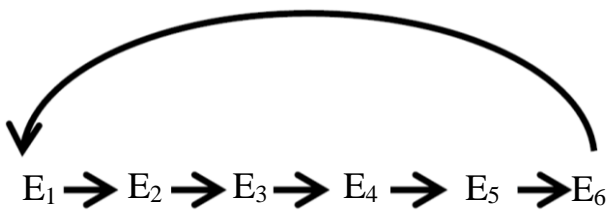

**Fig. 6: B** model of the turnstile

## Interview States Re-Modelled

Bochmann (2015) presented a model of an interview process involving what he called "agents" and "actions." the model comprises three agents: The company director, a journalist (who is doing the interview to write an article about the company to his/her journal) and a secretary. The interview includes a welcome handshake, interview talk and goodbye, jointly performed by the director and the journalist, as well as talk with the secretary and secretary entering or *leaving* jointly performed by the director and the secretary. The interview is interrupted when the secretary enters the room until he or she leaves. A sequencing of these actions is defined by the state diagram Fig. 7, which also represents the order of actions of the company director.

Figure 8 shows the **S** model of the interview example. The manager and the journalist enter the area of the interview (circles 1 and 2) where they engage in a handshake, (3) then interact with each other in the interview (4 and 5). The secretary enters the area (6) and he or she is joined by the director (7), interrupting the interview. The director and the secretary engage in conversation (8). Then the director returns to the interview (9) and the secretary exits (10).

The **S** description of Fig. 8 represents a typical interaction process. The director, journalist and secretary appear in the domain of the model along with the processes of handshaking, the interview and the conversation between the director and the secretary. Such a picture is atemporal because it places different scenes within the interview process side by side. The different scenes can be viewed as *changes* in the sense that each scene plays its role in the "theater" of the **S** plot. **S** does not exist as a *real* system, but it does embed potentialities of real scenes interwoven together without any time order. This does not imply the absence of the structure of sequentiality, because the flow (arrows) indicates a type of before and after relationship. The relationship is atemporal (similar to relationships such as "3, 4, 5" and "point, line, square," which embed some atemporal ordering; e.g., placing 3 before 4 has nothing to do with time).

*The D Model*

To simplify the turnstile example, we skipped important semantics related to the divisions of **S** to produce **D** as *events*. However, in the interview example, we have to tie the static (timeless) **S** to unfolding scenes, which we call *changes*, to (time) events.

We can view **S** as a collection of small scripts and the modeler's role is to identify consistent long scripts that may be constructed from the collection of short scripts. This can be achieved by slicing **S** into specific *changes* (identifying processes), hence producing the **D** model. The changes are timeless because **S** is atemporal. They emerge from the slicing of **S** and constitute different subdiagrams (machines/things). Each slice (change) is a scene that has its position in the series of changes, thus creating a chronology of changes by its relative position, not by its time of creation. A *change* here means *variation* in the sense of a difference from the rest of the furniture (all things occupying **S**) in the modelled domain (e.g., slices of an apple produced by a multicutter are borne at the same instance but different from each other in position). Modeling discourse requires this variation to establish pieces that allow the modeler to go beyond the static **S** and weave the model from these changes. Note that changes are state*s* that are produced in a systematic way. A state is one scene in **S**.

Accordingly, 10 changes (states) are identified in the **S** model of the interview example Fig. 9:

1. The director and the journalist enter the meeting room.
2. They shake hands.
3. They start the interview.
4. They process the interview.
5. The secretary enters.
6. The director leaves the interview to meet the secretary.
7. The director and the secretary converse.
8. The director returns to the interview.
9. The secretary leaves.
10. The director and the journalist end the interview.

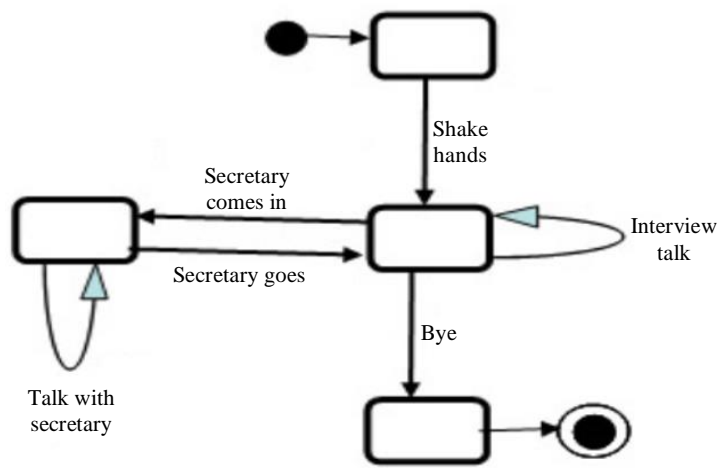

**Fig. 7:** Model of an interview (adapted from Bochmann, 2015)

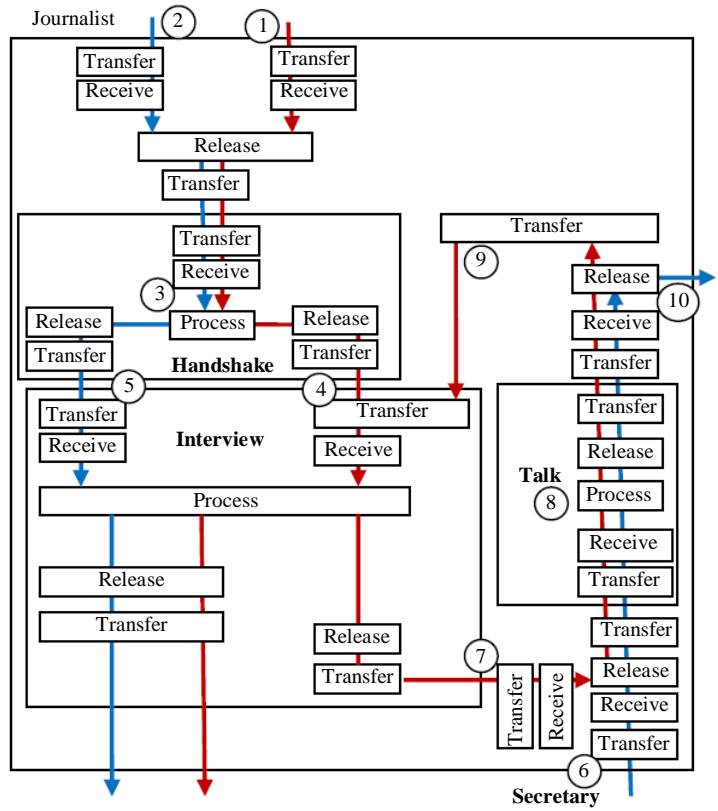

**Fig. 8:** S model of the interview

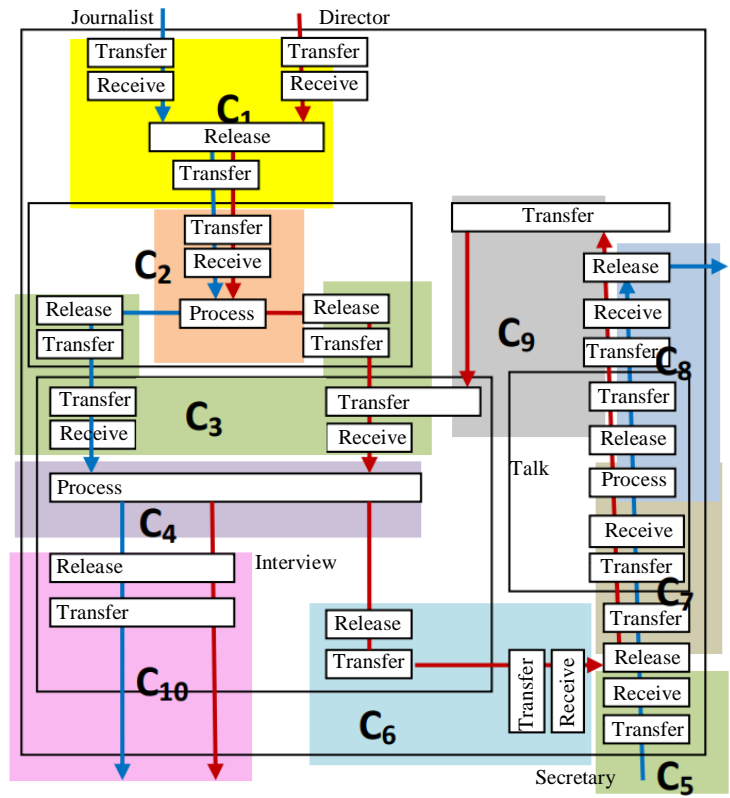

**Fig. 9:** D model of the interview

The TM flow in **S** dictates the order of changes: C1 → C2, then C2 → C3, then C3 → C4, then C4 → C5 OR C4 → C5, etc. This produces the chronology of changes shown in Fig. 10. Accordingly, we cannot view **S** as either a space or a time, but rather a frame for changes relative to each other; thus, order can be imposed (i.e., "before," "after," and "simultaneously"). It is possible that **S** includes several independent components of the chronology of changes that are not connected by flows or triggering. We assume that **S** has only one such chronology of changes (a graph with only one component).

Philosophically, the chronology of changes/states looks like a picture of the Aristotelian notion of *what things should be* (form). If the journalist and director meet (coming in), it is expected that they shake hands and then start the interview. If the secretary's interruptions are permitted, then it is expected that the director will talk with her and then return to the interview. The order of actions is a structural order, not a temporal order. The scenario is a template (form) of potential series of states/changes in the world. Accordingly, the chronology of Fig. 10 is a timeless order that reflects the rules governing the order of elements in TM actions.

*Time*

Figure 10 (changes and their chronology) reflects, in philosophy, the so-called B-series (of time), which is the series of all changes ordered in terms of logical relations such as "earlier than," "simultaneously," and "later than." The **S** structure covers multiple epochs of change that encroach on each other. At this point, after identifying all changes, we can insert time in the model, thereby creating events. Changes are potential (physical) events. For every event there is an accompanying change (a change is called a *fact* in philosophy, but philosophical facts *may* include time, thus mixing up time and events). Time is necessary to exclude an unreasonable sequence or duration of changes. Figure 11 shows the chronology of events in the interview example.

Thus, time is a mechanism that realizes (makes practical) plots of events. In our example, time physically realizes the changes in terms of "adequate starts and durations" (e.g., the interview occurs within an acceptable period instead of over many days; Fig. 12. The "after" is a relation between changes, but an "acceptable period" is a time-based imposition projected over the relation (e.g., the interview slice comes after the handshaking slice by minutes but not after, say, a day). Time brings practicality to changes in the model. Hence, the mere insertion of time in a change is a physical event specifying the start and duration of each change to make them fit together as a physical realization.

Additionally, with the introduction of time in the chronology of events, we can invent events such as the start time and end time for the whole chronology of events. Note that we cannot create a "start change" because all changes start simultaneously; hence, the *start change* precedes all changes. If we cut a watermelon in two slices, then both pieces are created simultaneously. However, we can create a start event.

Assume that Δ is a basic thimac (i.e., a thimac with no subthimac); for example, a single action such as create, process, release, transfer, or receive. Let ▲ denote the time thimacs of basic thimacs. Assume that σ is the smallest unit of time taken by any ▲. Accordingly, the σs are mapped to the positive integers:

$$\sigma 1, \sigma 2, \sigma 3, \ldots$$

We will assume that each event is mapped to *consecutive* σs that represent its start, duration and end time. The aim is to align events with respect to each other such that each occupies one or more σ. Any event will require multiple consecutive *σs*. We can conclude that an event has a *unique* time (hence, an event is unique) and a change may have different events and hence different times (different events over the same change).

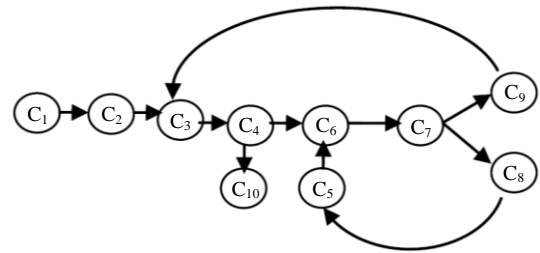

**Fig. 10:** Chronology of changes in the interview example

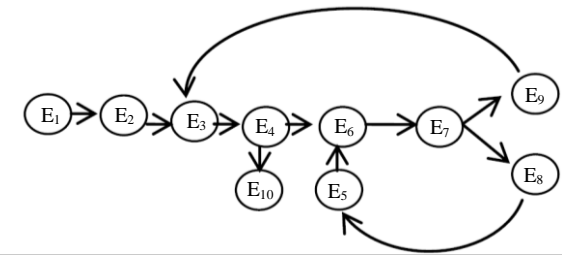

**Fig. 11:** Chronology of events in the interview example

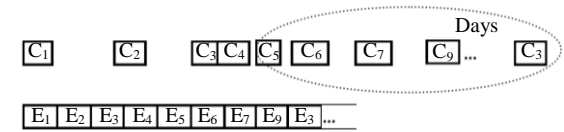

**Fig. 12:** Differences of realization of chronologies of changes (top) and events (bottom; e.g., The director leaves the interview to talk to the secretary, then comes back to the interview is an acceptable model of changes, but not of events.)

It is obvious that events are never repeated. Repeatability is a shorthand notation for events that have identical changes. Thus, $E_3$ (interviewing) denotes a set of events $E_3^1, E_3^2,\ldots, E_3^n$ such that $C(E_3^1) = C(E_3^2) \ldots = C(E_3^n)$. Saying "the sun rises every day" is a way of saying "the sun rose today, yesterday, before, or after," etc. The different events have the same change (sun rising) and differ only in their time subthimac.

A repeated single type of event (e.g., two knocks on a door) indicates two consecutive events with the same change. Events can be constructed from small consecutive events. In general, if $E_i$ happens before $E_j$, then it is not possible that $E_i$ follows $E_j$.

Changes also cannot be repeated except through events. The arrows in the chronology of changes indicate flow of things (note the difference in the type of arrows between Fig. 10 of changes and Fig. 11 of events). An apple that is decaying is a single change and a single event involving flow of decay (e.g., creating more bacteria), not a repeated change or event of decay. Thus, $C_9 \to C_3$ does not mean a repeat of $C_3$ but the *flow* of the director back to $C_3$. $E_9 \to E_3$ means the occurrence of event of *type* $E_3$ (has the same change as a previous event of type $E_3$).

Re-examining the interview example, Bochmann (2015) described the model in Fig. 7 as a Labelled Transition System (LTS). An LTS contains a finite number of states and a finite number of transitions that are labelled. Each label represents a certain interaction of the modeled object with its environment and the transition diagram defines the order in which these interactions may be performed. LTSs seem to be an active area of research. A very recent work involves LTS decompositions and their solvability by Petri nets (Devillers *et al*., 2019). According to (Fares *et al*., 2013), formal specification languages introduce entities, usually called *processes*, which offer similar operators and define their operational semantics based on an LTS.

The aim of re-modeling this example is to develop a better understanding of the formal LTS approach through TM modeling. However, the result provided an opportunity to exhibit some of the TM features, especially those that correspond to state-based modeling.

For example, in the LTS diagram Fig. 7, the secretary comes in then to reach the state where the director and the secretary engage in talk and this *talk* seems to be repeated (reflexive arrow). This repeatability of the state of talk between the director and the secretary, according to the TM model, is flow-based repetition ($C_9$ to $C_3$ and $C_9$ and $C_5$ in Fig. 10). According to (Bochmann, 2015), "In this example (and when using LTS models, in general), one uses a state-oriented modeling approach where the actions are modeled as transitions and not as 'activities' or 'states'. In fact, these transitions here represent collaborations between several agents."

A transition is represented as an arrow that indicates what activity can be executed next. A transition is triggered by an event. There is no direct definition of what an event is in LTS. In LTS, kinds of events are given as messages, changes of the values of variables and events that trigger time (e.g., timeout). Therefore, to understand an event, we have to look at what a message is, what a change is and what triggering time means.

Returning to the repeatability of talk with the secretary in Fig. 7, it seems to involve time, but the window of time is separate. Accordingly, "talking to the secretary" involves many transitions that lead to the same state. However, transitions are triggered by events. At this point, it is not possible to understand the semantics of this part of the LTS diagram, especially regarding the timing aspects. A similar statement applies to the reflexive arrow over the interview.

On the other hand, the TM modeling seems to provide a simple description of the interviewing process as follows:

1. Description in terms of the S model with atemporal flows specified by five actions
2. The **D** model identifies the states or changes. Hence, we can construct the atemporal chronology of states
3. The behavior model model applies time to states to build the temporal chronology of events

Each of the notions of a thimac, machine, thing, action, flow, change or state, event and time has its own precise definition and ways of participating in the models. For example, an event is a change plus time. A change is a scene as a subdiagram of S. The subdiagram of S is constructed from TM actions, which are create, process, release, transfer and receive.

## A Traffic Light Control Re-modeled

This section applies TM modeling to a larger project involving engineering design, which brings the analysis closer to the typical state machines in the literature.

Wagner *et al*. (2006) designed a traffic light control at a level crossing of a railway and a road. In this example, there is only one rail line, but the trains may come from either direction. Three sensors detect the trains: L (left), M (middle) and R (right). The output of the system is a red lamp that should be switched on if the train approaches sensor L (if coming from left) and should be switched off if the train leaves the sensor at *M*. Similar actions should be performed if the train comes from the right. The state diagram Fig. 13 to model the system is described in Table 1, where *X* is either *L* or *R*. (Wagner *et al*., 2006) stated that using state diagrams is a way to rethink design software, in which most of the usual control-flow coding is avoided and the FSM concept takes a predominant part in the design process.

According to (Wagner *et al*., 2006), FSM introduces "a concept of a state as *information about its past history*" (Italics added). Other mentions of the notion of state include "The history of input changes required for clear determination of the state machine behavior is stored in *an internal variable* State" (Wagner *et al*., 2006). In the given example, the each state is clearly identified.

Figure 14 shows the S model of this railway example. The train comes from the left (1) to enter the left-middle area (2), triggering the red light (3 and 4). It continues to the middle area (5), then to the middle-right area (6) to trigger the green light (7 and 8). Then, the train leaves (9) to the right.

Additionally, the train may come from the right area (10) to the middle right area to trigger the red light (11 and 12). It continues to the middle area (13), then to the middle-left area (14) to trigger the green light (15 and 16). It exits to the left area (16).

Figure 15 shows the D model. The changes in the figure are identified according to (Wagner *et al*., 2006) given states as follows:

- Change 1 ($C_1$): The train is at the L sensor. This change is captured by the L sensor when the transfer of the train starts from the L to the L-M area, as shown in C1 in Figure 15. This is what (Wagner *et al*., 2006) call state 001 Table 1
- Change 2 (C2): Now the train is completely in the L-M area. The L sensor now creates a signal to turn the light red, which (Wagner *et al*., 2006) call state 011 Table 1
- Change 3 (C3): The train is at the M sensor. This is what (Wagner *et al*., 2006) call state 011 Table 1. It is not clear what the role of the sensor is, except perhaps communicating the value 011

- Change 4 ($C_4$): The train moves to the M-R area to trigger turning the light green and continues to the R area. This is what (Wagner *et al*., 2006) call state 100 Table 1

Similarly, changes $C_5$, $C_6$, $C_7$ and $C_8$ occur when the train comes from the right. Note that:

- $C_1$ and $C_5$ are of state type 001
- $C_2$ and $C_6$ are of state type 010
- $C_3$ and $C_7$ are of state type 011
- $C_4$ and $C_8$ are of state type 100
- $C_5$ and $C_{10}$ are of state type 101

Note also the roles of the *L* sensor:

(a) Recognizing that the train is coming from the left
(b) Sending a signal to turn the light red

The R sensor's roles are as follows:

(a) Send a signal to turn the light green
(b) Signal that the train is going to the left

The R and L sensors play similar roles.

**Table 1:** State model of the train example (adapted from Wagner *et al*., 2006)

| Description | State | Code |
|---|---|---|
| No train | No train | 000 |
| On X | Coming | 001 |
| Between X and M | Approaching | 010 |
| On M | Present | 011 |
| Between M and X | Leaving | 100 |
| On X | Going | 101 |

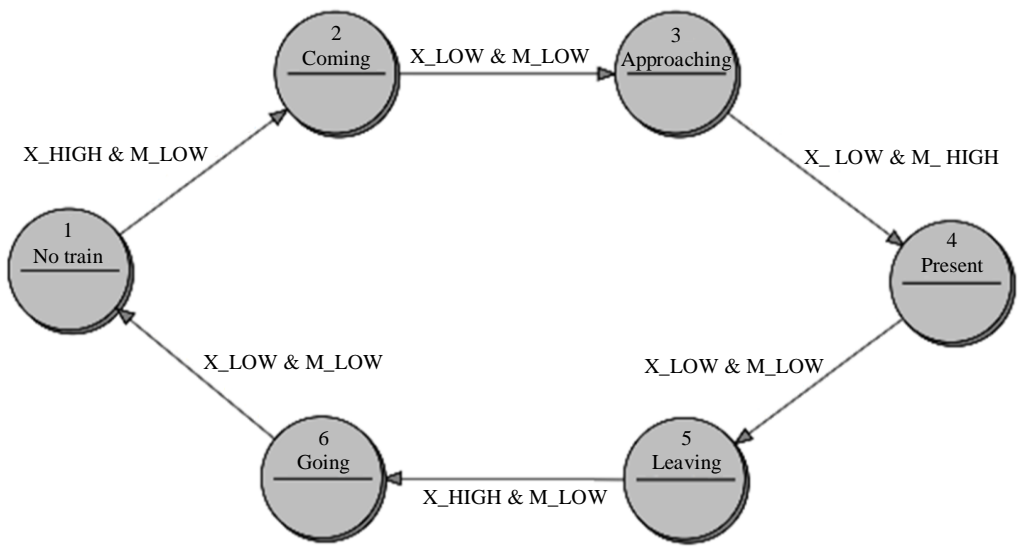

**Fig. 13:** State model of the train example (adapted from Wagner *et al*., 2006)

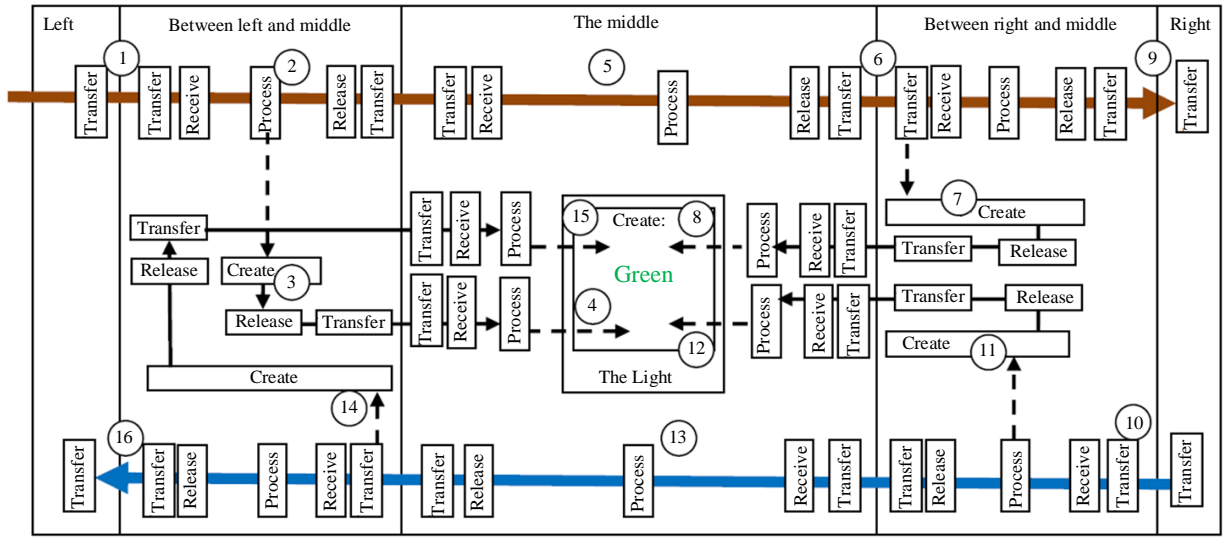

**Fig. 14:** S model of the railway crossing

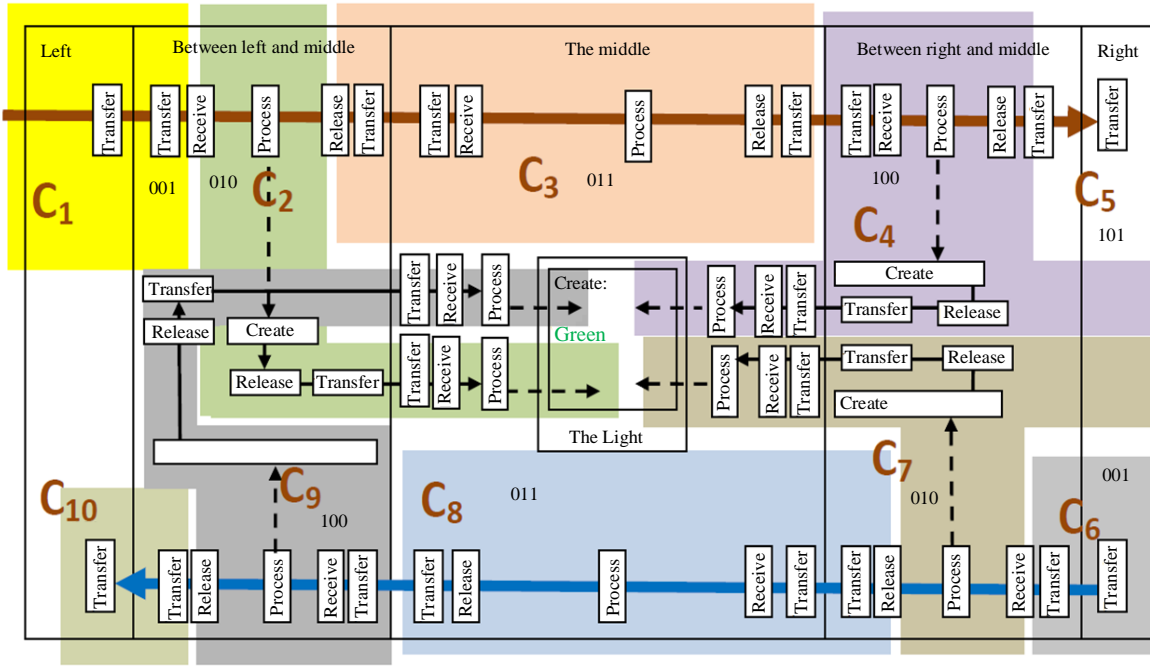

**Fig. 15:** D model of the railway

From the TM perspective, the locations of the sensors are problematic in (Wagner *et al.*, 2006) description of the problem. For example, according to the TM model, the L sensor should be located at the end of the L-M area (transfer from L to L-M) *furthest from the M area*. Thus, the sensor functions as follows:

1. As soon as the front of the train enters (transfer) the L-M area, the sensor recognizes the train is "On L" Table 1

2. As soon as the end of the train enters the L-M area, the sensor signals "between *L* and *M*"

However, when the train comes from the right, leaving the M area and entering the M-L area, the sensor should be located at the end of the M-L area *closest to the M area*.

An additional observation of the FSM modeling of the train example is that when the states are converted to events, care should be taken in possible conflicts in behaviors (or sub-behaviors) of the system. For example, in (Wagner *et al.*, 2006) state machine, to avoid allowing trains to come from the left and right simultaneously, they equated states $C_0$ (train from left) with $C_6$ (train from right) and made them one state. Because only one state is active, no conflict occurs. However, the orientation (direction) of the two is different so the two states are not identical.

We will not elaborate further on such problems, focusing instead on contrasting the semantics of FSM in terms of the TM modeling.

Figure 16 shows the succession of changes. As stated previously, generating changes is analogous to cutting **S** into pieces in the sense that the changes are born simultaneously. The cutting is a timeless process. When we use a multicutter, say for an apple, then all pieces stay in their positions relative to each other after the cutting process. Each piece is a *change* in the total apple. The slices' positions relative to each other may be taken as a base to order them and we call the succession a chronology of changes. The whole order is timeless: A change is "before" or "after" another change only in the selected ordering. This is what happens when we construct the D model. In the chronology of changes in Fig. 16, we note that we can start the chronology (timeless ordering) of changes at $C_1$ or $C_6$. Note that if we introduce a "no train" event, say, $C_0$ = "no train" (as Wagner *et al.*, 2006) as a change (state), then $C_0$ is adjacent to all changes because they are created simultaneously. This would disturb the diagram of the chronology of changes.

Again, the chronology of states (e.g., state machine) creates a picture of what things should be (form). If a train is coming, then things progress as follow: The train crosses the area before arrival and then it arrives. Next, it moves away. The order is a structural not temporal. It is a template (form) of a potential series of motions in the world. Accordingly, the actions or stages in the TM machine Fig. 1 are just changes or states and their chronology is a timeless order that reflects laws of order over elementary states. Of course, it is possible that a thing flows from release back to process, but this is not what things should be.

To develop the behavior of the system, all changes are converted to events Fig. 17. The events diagram can be justified as a construction process that involves time.

In the analogy mentioned previously, suppose that we want to cut the apple into 10 pieces. We first cut one piece, then cut a second piece from the rest of the apple and so on. Although the result is identical to the simultaneous multicutting process, each piece now has a temporal stamp that is either born before or after. Because we are in the time context, we can declare the starting time of the cutting process as Event 0 ($E_0$). $E_0$ is followed exclusively by $E_1$ or $E_6$. Thus, we arrive at the behavior of the system shown in Fig. 17. Now, $E_1$ and $E_6$ cannot occur simultaneously. Wagner *et al.* (2006) "no train" event can be amended to the end or beginning of the train flows.

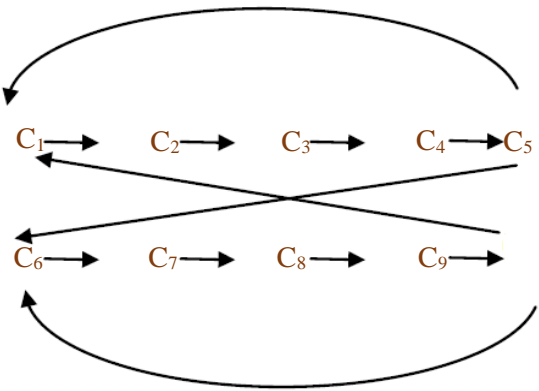

**Fig. 16:** Chronology of changes in the railway example

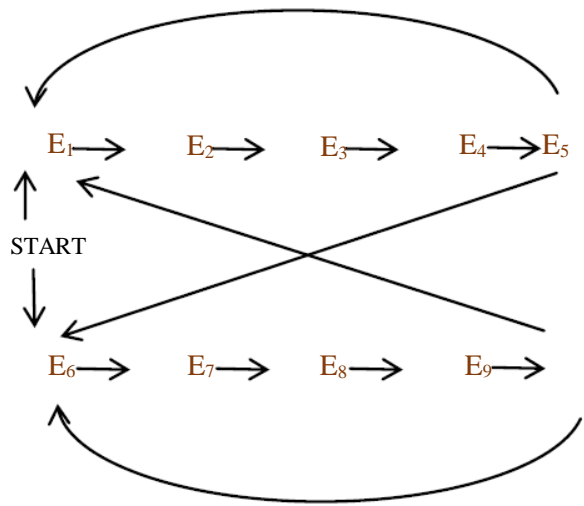

**Fig. 17: B** model of the railway

Accordingly, we can declare that a state is a (timeless) change. The change is a discrete snapshot of a portion (subdiagram) of the grand thimac. Because it is a change, the notion of state is a pre-events snapshot. The change or state is a sub-thimac and the order of changes or states does not need the notion of time. For example, the state "On M" in the train example is a thimac and its machine is:

*Middle. Transfer (input).Receive.Process.Release. Transfer (output)*

That is, the corresponding machine is the flow of the train inside the middle area. "On M" is just a name for the thimac. Hence, each of the five actions in M can represent an elementary state. Accordingly, the TM Fig. 1 is a state machine with five elementary states.

## Conclusion

This paper aims to establish a precise definition of the notion of state and state machines. State is the main notion of a state machine, in which events drive state changes. The analysis of these concepts is based on a new modeling methodology called the Thinging Machine (TM) and we considered a number of examples of existing models.

The TM model seems to provide richer descriptions of the situations in the given examples with clear meaning of what a state (change) is. It can be used as a semantic base for the state diagram. The state machine is obviously a thimac without time and the order of the states is a timeless order. This order is isomorphic to the time order of states with time.

## Ethics

This article is original and contains unpublished material. No ethical issues were involved and the author has no conflict of interest to disclose.